\documentclass[useAMS,usenatbib,usedcolumn]{mn2e}

\usepackage[utf8]{inputenc} 
\usepackage{Sweave}
\usepackage[OT1]{fontenc}
\usepackage{hyperref}
\usepackage{graphicx}
\usepackage{bm}
\usepackage{amsmath,amssymb}
\usepackage{aas_macros}

\usepackage{color}

\newcommand{\rvir}{r_{\rm vir}}

\newcommand{\hmsun}{h^{-1}M_{\odot}}

\newcommand{\eproj}{\mathsf{e}}
\newcommand{\eplus}{\eproj_{+}}
\newcommand{\ecross}{\eproj_{\times}}
\newcommand{\emag}{|\eproj|}

\begin{document}

\bibliographystyle{mn2e}

\setkeys{Gin}{width=0.5\textwidth}
  
\title[GAMA Radial Alignments]{Galaxy and Mass Assembly (GAMA): galaxy radial alignments in GAMA groups}

\author[ M. D. Schneider et al.]{Michael D. Schneider$^{1,2}$\thanks{E-mail: schneider@ucdavis.edu}, 
Shaun Cole$^{3}$, Carlos S. Frenk$^{3}$,
Lee Kelvin$^{4, 5}$, 
\newauthor
Rachel Mandelbaum$^{6,7}$, 
Peder Norberg$^{3}$,
Joss Bland-Hawthorn$^{8}$,
Sarah Brough$^{9}$,
\newauthor
Simon Driver$^{4, 5}$,
Andrew Hopkins$^{9}$,
Jochen Liske$^{10}$,
Jon Loveday$^{11}$,
Aaron Robotham$^{4, 5}$\\
$^{1}$Department of Physics, University of California, One Shields Avenue, Davis, CA 95616, USA.\\
$^{2}$Lawrence Livermore National Laboratory, P.O. Box 808 L-210, Livermore, CA 94551-0808, USA.\\
$^{3}$Institute for Computational Cosmology, Department of Physics, Durham University, South Road, Durham, DH1 3LE, UK.\\
$^{4}$School of Physics and Astronomy, University of St Andrews, North Haugh, St. Andrews, Fife KY16 9SS.\\
$^{5}$International Centre for Radio Astronomy Research, 7 Fairway, The University of Western Australia, Crawley, Perth, WA 6009, Australia.\\
$^{6}$Princeton University Observatory, Peyton Hall, Princeton, NJ 08544 USA.\\
$^{7}$Department of Physics, Carnegie Mellon University, Pittsburgh, PA 15213, USA.\\
$^{8}$Sydney Institute for Astronomy, University of Sydney, NSW 2006, Australia. \\
$^{9}$Australian Astronomical Observatory, PO Box 296, Epping, NSW 1710, Australia. \\
$^{10}$European Southern Observatory, Karl-Schwarzschild-Str. 2, 85748 Garching bei M\:{u}nchen, Germany.\\
$^{11}$Astronomy Centre, University of Sussex, Falmer, Brighton, BN1 9QH, UK.\\
}

\date{
LLNL-JRNL-544911}


\pagerange{\pageref{firstpage}--\pageref{lastpage}} \pubyear{2012}

\maketitle

\label{firstpage}

\begin{abstract}
We constrain the distributions of projected radial alignment angles of satellite 
galaxy shapes within the Galaxy And Mass Assembly survey group catalogue.  
We identify the galaxy groups using spectroscopic redshifts and measure galaxy 
projected ellipticities from Sloan Digital Sky Survey imaging. 
With a sample of 3850 groups with 13655 satellite 
galaxies with 
high quality shape measurements, 
we find a less than 2-$\sigma$ signal of radial alignments in the mean projected ellipticity 
components and the projected position angle when using galaxy shape estimates optimized 
for weak lensing measurements.
Our radial alignment measurement increases to greater than 3-$\sigma$ 
significance relative to the expectation for no alignments 
if we use 2-D S\'{e}rsic model fits to define galaxy orientations.
Our weak measurement of radial alignments is in conflict with predictions 
from dark matter $N$-body simulations, which we interpret as evidence for 
large mis-alignments of baryons and dark matter in group and cluster satellites.
Within our uncertainties, that are dominated by our small sample size, 
we find only weak and marginally significant trends of the radial alignment angle 
distributions on projected distance from the group centre, host halo mass, and redshift 
that could be consistent with a tidal torquing mechanism for radial alignments. 
Using our lensing optimized shape estimators, we estimate 
that intrinsic alignments of galaxy group members may contribute a systematic error to  
the mean differential projected surface mass density of groups 
inferred from weak lensing observations by 
$-1\pm 20$\% at scales around 300~$h^{-1}$kpc from the group centre assuming a
photometric redshift r.m.s. error of 10\%, and given our group sample
with median redshift of 0.17 and median virial masses $\sim10^{13}$~$h^{-1}M_{\odot}$.
\end{abstract}


\section{Introduction}
\label{sec:introduction}
The hierarchical model for cosmological structure formation posits that groups and 
clusters of galaxies form by the accretion of smaller groups and individual galaxies.  
In this scenario, as galaxies are accreted into a group they would be tidally torqued 
so that their major axes would be aligned with the centre of the gravitational 
potential well.  The efficiency of torquing within a cluster 
should depend on the gradient of the potential well 
and the eccentricity of the accreted galaxy's orbit~\citep{pereira08} as well 
as the rotational support (or lack thereof) of the infalling galaxy~\citep{wesson84}.  
This simple picture then predicts 
that satellite galaxies should be more radially aligned in more concentrated 
(i.e. typically lower mass) groups and that 
the degree of alignment should have an inverse relation to the angular speed of the galaxy 
thereby imparting a dependence on the distance from the potential centre~\citep{kuhlen07, pereira08, pereira10}.

This tidal torquing model for the radial alignments of satellite galaxies could be complicated by 
effects such as the tidal stripping of infalling satellites or the misalignments of stars 
and dark matter due to complex accretion and merger histories. 
Alternatively, if the time-scales for tidal torquing in a group or cluster are comparable 
to the age of the Universe, then any global alignments of cluster galaxies may instead serve as 
a probe of the anisotropic accretion history from filaments around the 
cluster~\citep{djorgovski83, wesson84, adami09, song12}, 
in which case no radial alignments should be detected.
Observations seeking to measure radial alignments are further confounded by the 
difficulty in measuring unbiased galaxy shapes and orientations, the
projection of unknown 3D galaxy morphologies into the plane of the sky,
and the unknown location of the group or cluster potential centre.

Using photographic plates covering three nearby clusters, 
\citet{hawley75} rejected the null hypothesis of uniform projected radial 
alignment angles at roughly 98 percent significance in the Coma cluster (only).
\citet{djorgovski83} later found significant radial alignments in the Coma cluster, with 
faint and red galaxies near the cluster centre showing the strongest alignments.
With the large galaxy samples in the more recent redshift and cluster surveys, several groups have 
claimed both strong detections of radial alignments~\citep{pereira05, agustsson06, faltenbacher07}
and null detections~\citep{bernstein02, siverd09, hao11, hung12, blazek12}.
See~\citet{hao11} their table 1, for a comparison of measurements in the Two-degree-Field Galaxy 
Redshift Survey (2dFGRS) and Sloan Digital Sky Survey (SDSS).
\citet{siverd09} and \citet{hao11} showed that measurements of satellite galaxy 
orientations using SDSS isophotes may be subject to numerous systematics that could 
potentially resolve the discrepant claims of detection and null signals in the literature.
We will address this issue further in Section~\ref{sec:isophoteComparison}.

The 3D radial alignments of dark matter sub-haloes within cluster-sized parent haloes 
have been measured in $N$-body simulations~\citep{kuhlen07, pereira08, knebe08}, 
showing much stronger radial alignments than in any observations independent of parent or 
sub-halo mass.
\citet{kuhlen07} and \citet{knebe08b} showed that 
using only the inner 10--20\% of the particles in a dark matter sub-halo, rather than 
all bound sub-halo particles, introduces significant scatter in the distribution of radial 
alignment angles that is more consistent with previous observations.
Using $N$-body simulations with gas and star formation physics included, \citet{knebe10}
concluded that gas physics does not measurably affect the radial alignment angles of 
satellite galaxies relative to the orientations inferred from studying the parent 
dark matter sub-haloes alone.
\citet{pereira10} performed isolated $N$-body simulations of a satellite falling into a cluster potential 
to study the tidal torquing effect as a function of orbital phase and find a strong dependence 
of the radial alignment angle on the orbital angular velocity, which was a conclusion 
also found by \citet{kuhlen07}~and~\citet{knebe10} in their simulations embedded in a cosmological environment.
Together, these simulation studies have established the tidal torquing mechanism 
as the dominant effect on satellite radial alignments and as a key prediction for 
structure formation in cold dark matter theories.

The anisotropic accretion of satellites on to groups and clusters also tends to align
satellites with the parent cluster major axis~\citep[e.g.][]{dekel85, plionis03}.
The measurement of this effect is beyond the scope of this paper, but we note that 
such alignments could dilute the radial alignment measurement when approximating 
triaxial groups with a spherical geometry.


Models for the alignments of group and cluster members are also important for 
predicting and mitigating the intrinsic alignment contamination in weak lensing 
studies~\citep[e.g.][]{hirata04b, mandelbaum06, hui08, schneider10, kirk10}. 
\citet{bernstein02} explicitly constrained the lensing contamination from their 
measured radial alignments, but we will focus more on constraining models for the
radial alignment angle distributions that could be later propagated into 
predictions for intrinsic alignment contamination in lensing measurements.

In this paper we constrain the distributions of projected radial alignments of galaxy 
group satellites in the 
Galaxy and Mass Assembly (GAMA) 
survey\footnote{\url{http://www.gama-survey.org/}}~\citep{driver09, baldry10, robotham10, hill11, driver11}. 
All group members in the GAMA catalogue are spectroscopically confirmed and 
the group properties have been calibrated by comparison with mocks built 
from $N$-body simulations~\citep{robotham11}.
As in some other recent studies, we measure satellite galaxy orientations 
from SDSS imaging. We use two estimates of the galaxy shapes, the 2D galaxy model fits to 
SDSS $r$-band imaging data which account for the effects of point spread function (PSF) convolution using the SIGMA 
pipeline as described in~\citet{kelvin12} 
and a galaxy shape estimator optimized for weak lensing that is 
more sensitive to the shapes of galaxies at smaller radii~\citep{hirata03, mandelbaum05}.
 
This paper is organized as follows. 
We describe the relevant features of the GAMA galaxy group catalogue in Section~\ref{sec:groupcatalogue}
and the galaxy shape estimators in Section~\ref{sec:galaxyshapes}. 
We present and compare our measures of galaxy radial alignments in groups in 
Section~\ref{sec:alignmentmeasures}. We use mock group catalogues with model radial 
alignments as described in Section~\ref{sec:mocks} to assess the significance of our measurements.
We then show our measurements in Section~\ref{sec:results} and discuss their 
implications in Section~\ref{sec:conclusions}.
To aid the comparison with previous analyses based on SDSS imaging, we compare 
our shape estimators with the isophote measurements in the 
SDSS Catalog Archive 
Server\footnote{\url{http://cas.sdss.org/}}~\citep[{\tt CAS},][]{sdsscas} in Appendix~\ref{sec:isophoteComparison}.

\section{Data description}
\label{sec:data}
In this section we describe the data sets and analysis pipelines 
we have combined to perform our analysis.

\subsection{Group catalogue}
\label{sec:groupcatalogue}
We use the GAMA-I galaxy group catalogue (G$^3$Cv1) 
as described in \citet{robotham11} to define the group memberships and global group properties.
The G$^3$Cv1 catalogue contains 4263 groups 
with three or more members identified within $\sim$142~deg$^2$ 
with a spectroscopic depth
limit of $r_{AB}=19.4$ 
\citep[with 98\% completeness][]{driver11}.
We define group redshifts to be the 
median redshift of all group members, which span 
0.017 to 0.46
with a mean redshift for the group catalogue of 0.18.
By comparing with mock group catalogues built with $N$-body simulations, \citet{robotham11} 
assigned dark matter halo masses to each group. 
We assign halo masses 
to match the observed 
group velocity dispersions, which can lead to some spuriously large or small halo mass estimates. 
However, 95 percent of the groups with three or more members have halo masses in the range 
$4.4\times 10^{10}$ to 
$8.6\times 10^{14}$~$\hmsun$, 
where $h$ is reduced Hubble constant.
In addition to the robust determination of group membership,
our study is sensitive to the determination of the group centre about which the satellite 
galaxy alignments are measured.
Throughout, we use the ``iterative group centre'' that \citet{robotham11}
showed to be more accurate than assuming the Brightest Cluster Galaxy 
(BCG) is at the group centre (if the BCG can be 
accurately determined).

The observed angular separations between satellite galaxies and the group centres are converted to 
physical comoving distances using the median group redshift to calculate the angular diameter 
distance to the group assuming a cosmology of $\Omega_{\rm m}=0.25$ and $\Omega_{\Lambda}=0.75$ (which 
is used consistently throughout the GAMA group catalogue construction).

\subsection{Galaxy shape measurements}
\label{sec:galaxyshapes}
For our primary method of galaxy shape determination, we use the shape measurements from the
REGLENS pipeline applied to SDSS imaging as described in \citet{hirata03} and \citet{mandelbaum05}.
The REGLENS shape estimates use a re-Gaussianization~\citep{hirata03} algorithm to correct for 
the effects of the PSF on the observed galaxy shapes. 
Briefly, REGLENS finds best-fit (in the least-squares sense) 2-D Gaussians to both 
the PSF and the observed galaxy image. The galaxy ellipticity is defined 
by a 2-D covariance matrix derived from the difference of the best-fit image and PSF 
covariances. A `resolution' factor is then defined by the fractional deviation 
of the traces of the PSF and the re-Gaussianized image covariances. 
First order corrections to both the resolution factor and ellipticity 
estimate are applied to account for deviations of the PSF and image from a Gaussian profile.
Because the Gaussian profile is a steeply falling function of galactic radius,
REGLENS tends to be most sensitive to the shapes of galaxies at much smaller radii than 
typical model fits, which we describe next. This is an advantage for the purpose of 
unbiased shear estimation in weak lensing studies, but may not be the optimal choice for 
probing the physical mechanisms behind intrinsic galaxy alignments.
The REGLENS pipeline also includes quality cuts based on the resolution of the galaxy images, 
galactic extinction, and seeing quality that reduce our group catalogue to
3862 groups with 13956 galaxies
from 4263 groups with 
21132 galaxies before the shape quality cuts are applied.
The size cuts for the REGLENS pipeline are described in 
section~2.2.1~of~\citet{mandelbaum05}. 
The REGLENS galaxies must have a resolution factor of $> 1/3$.
In addition, we select only those satellite galaxies with ellipticity magnitudes, as 
measured by the REGLENS pipeline, greater than 0.05. This minimum ellipticity magnitude 
cut further reduces our sample to 3850 groups and 
13,655 galaxies.

For comparison with previous studies of radial alignments and to test the 
robustness of our results with respect to the choice of galaxy shape 
estimator, we also define galaxy shapes
based on a 2-D S\'{e}rsic model fit to $r$-band SDSS imaging output by the 
SIGMA pipeline as part of the GAMA survey and described in~\citet{kelvin12}.
The SIGMA outputs can be found in the 
GAMA S\'{e}rsic Photometry catalogue, version 7 (SersicCatv07).
The SIGMA outputs we use here are the effective half-light radius along the 
semimajor axis $r_{\rm e}$, galaxy ellipticity $e_{\rm SIGMA}\equiv 1 - b/r_{\rm e}$ with $b$
the semiminor axis length, the 
position angle $\theta$ (relative to a fixed coordinate system) and 
S\'{e}rsic index (used in Section~\ref{sec:results} as a proxy for morphology). 
\citet{kelvin12} fit a PSF-convolved 2-D S\'{e}rsic model as well as 
neighbouring stars and galaxies so that the SIGMA outputs should not be 
strongly contaminated by the rounding effect of the PSF or 
blending with nearby objects. The PSF model is defined by centroid and width parameters
and is fit earlier in the SIGMA pipeline before fitting the galaxy profiles.
For consistency with REGLENS and common weak lensing analyses, we redefine 
the ellipticity magnitude in the SIGMA catalogue as,
\begin{equation}\label{eq:isoellipticity}
  e \equiv \frac{\mathtt{a}^2 - \mathtt{b}^2}{\mathtt{a}^2 + \mathtt{b}^2},
\end{equation}
where $a\equiv r_{\rm e}$ and $b\equiv r_{\rm e} (1 - e_{\rm SIGMA})$
We further define ellipticity measurement errors by 
formally propagating the reported SIGMA errors in $r_{\rm e}$ and $e_{\rm SIGMA}$.
We discarded approximately 2\% of the galaxies that 
passed the REGLENS quality cuts at this stage 
because the formally propagated ellipticity errors were outside the interval $[0,1]$.
In contrast to previous radial alignment measurements using SDSS isophotes, 
the ellipticities obtained from the SIGMA pipeline have
stellar PSF models incorporated in the fits to the galaxy profiles.
In Appendix~\ref{sec:isophoteComparison} 
we compare the SIGMA and 
independent GAMA isophote properties 
with the isophotes available in the SDSS {\tt CAS} catalogue, 
which lacks the PSF correction present in SIGMA.

The REGLENS shapes are weighted to measure the inner shapes of galaxies while 
the SIGMA shapes also utilize information on 
the shape in the far outskirts of a galaxy image.
The SIGMA shapes are potentially more sensitive to bias from sky background 
subtraction uncertainties and nearby neighbours in the imaging~\citep{siverd09}, 
although the careful pipeline in~\citet{kelvin12} attempts to mitigate these issues.
This also causes the SIGMA shapes to be more sensitive to astrophysical mechanisms 
that affect the outskirts of galaxy light distributions, which may include the 
tidal torquing mechanism we investigate here. 

\section{Methods}
\label{sec:methods}
\subsection{Measures of radial alignments}
\label{sec:alignmentmeasures}
We consider two common statistics for measuring the projected radial alignments of 
galaxies in groups and clusters.

\subsubsection{Position angle}
\label{sec:positionangle}

\begin{figure*}
  \centerline{
    \includegraphics[width=0.5\textwidth]{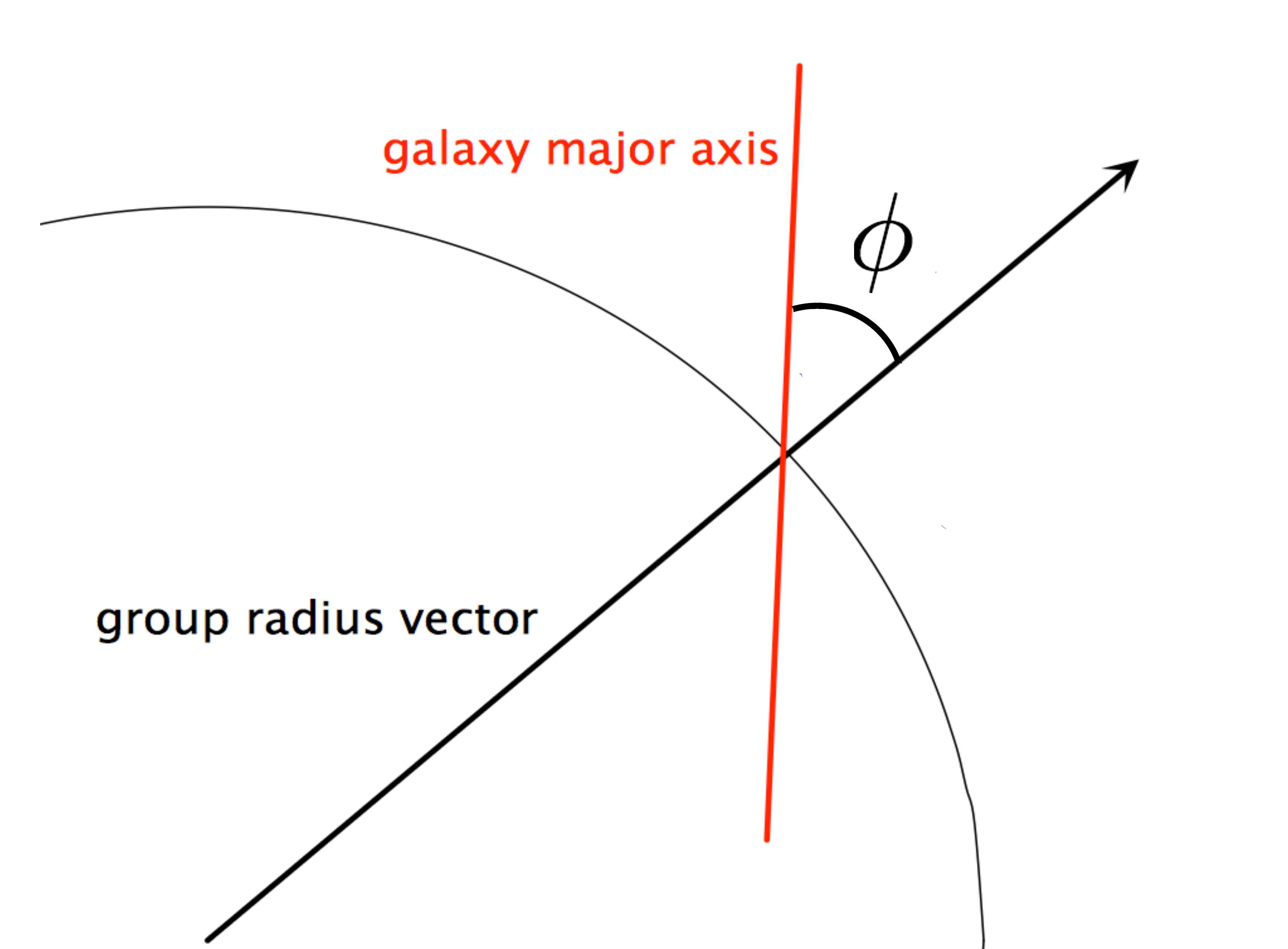}
    \includegraphics[width=0.5\textwidth]{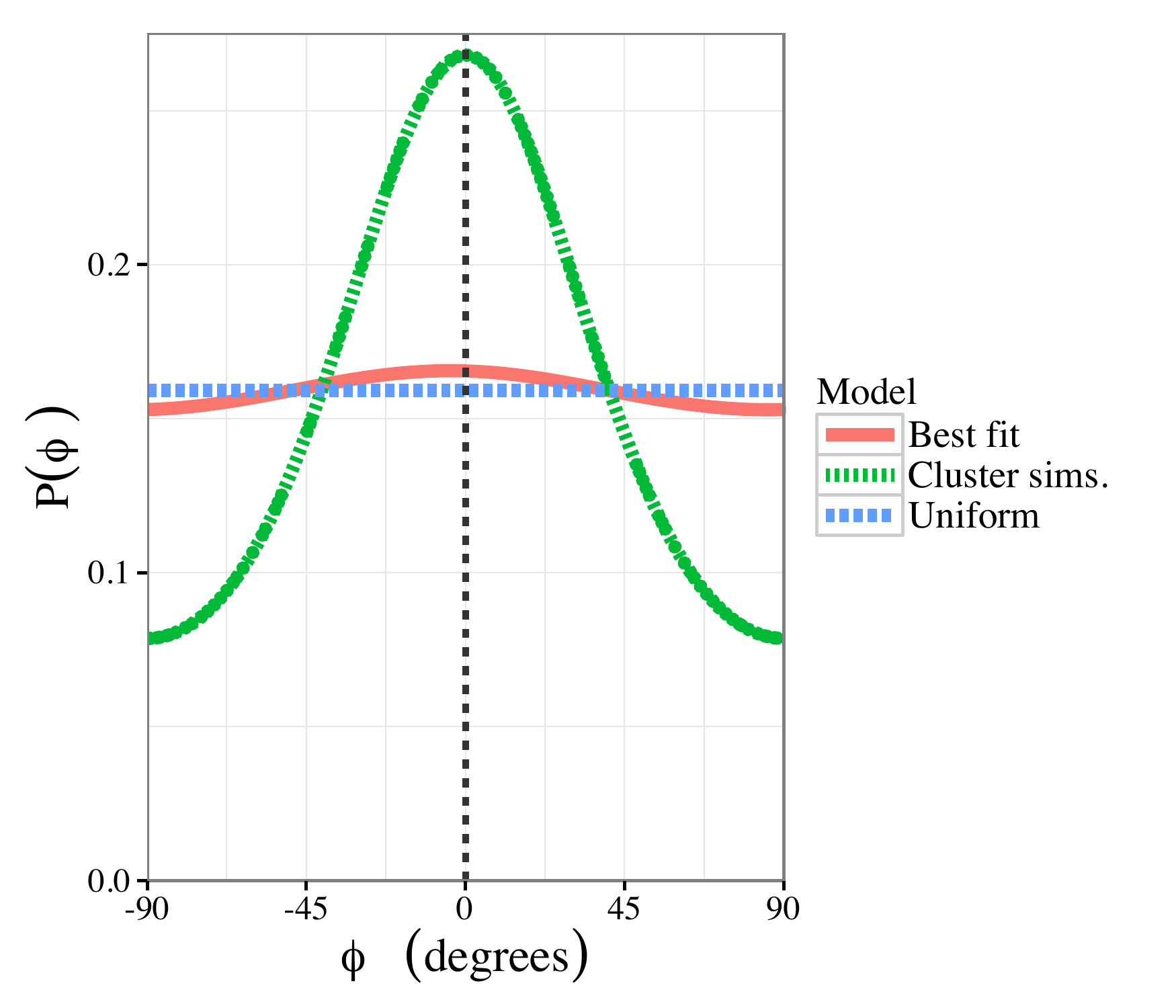}
  }
  \caption{Left: definition of the \emph{position angle} ($\phi$) of a satellite galaxy. 
  Due to the rotational symmetry in the plane of the sky, 
  the position angle is defined on $[-\pi/2, \pi/2]$.
  Right: the probability distribution of the position angle based on the von~Mises distribution (see the text) for 
  the best fit to the GAMA groups with three or more members (red solid line), a prediction from isolated cluster 
  $N$-body simulations (short dashed green line), and the uniform distribution (long dashed blue line).}
  \label{fig:positionangle}
\end{figure*}

The galaxy position angle, $\phi$, is defined as the angle between the projected 
galaxy shape major axis (however this is defined) and the projected radius vector of 
the galaxy position from the group centre, as shown in the left-hand panel of Fig.~\ref{fig:positionangle}.
When the position angle equals zero, the galaxy is perfectly aligned with the projected group 
radius vector. Because the galaxy shape is symmetric under 180$^{\circ}$ rotations in the plane of the sky,
the position angle is defined in the interval $(-\pi/2, \pi/2]$.
In the right-hand panel of Fig.~\ref{fig:positionangle} we show some model probability distributions for 
$\phi$, with periodicity imposed. The short-dashed green line is a fit to the distribution found 
in the $N$-body simulations of~\citet{pereira08} while the solid red line is the best fit to our 
data as described in Section~\ref{sec:results} below. The long-dashed blue line shows the uniform distribution.

\subsubsection{Ellipticity components}
\label{sec:ellipticitycomponents}
Given a measurement of the projected ellipticity components of a galaxy in a global 
coordinate system, the $x$ and $y$ components of the projected ellipticity can be 
conveniently represented as the real and imaginary parts of a complex number,
\begin{equation}
	\eproj = \emag\, e^{2i\phi_{e}}.
\end{equation}
For considering orientations within a group or cluster it is convenient to 
further define the rotated ellipticity components, 
\begin{equation}\label{eq:epluscross}
	\eplus + i\ecross = -\eproj\, e^{-2i\phi}, 
\end{equation} 
where $\phi$ is the azimuthal angle of the galaxy projected position with respect to the centre of the group, 
i.e. the position angle as defined in Section~\ref{sec:positionangle}.
A positive $\eplus$ component indicates a tangential alignment of the galaxy 
with respect to the group centre while a negative $\eplus$ indicates a radial alignment.
The $\ecross$ ellipticity component indicates satellite galaxy orientations at 
$\pm 45^{\circ}$ to the galaxy position vector. The $\ecross$ component 
is expected to have a zero mean for every group in the absence of a coherent ``curl'' component 
in the galaxy alignments, which is not motivated by any physical model that we know of.
Therefore, a group with radially aligned galaxies would have 
negative mean $\eplus$ and zero mean $\ecross$ components.

The observed ellipticity components are a combination of the intrinsic projected ellipticities 
of the galaxies and shears induced by gravitational lensing. 
Because the foreground lensing masses are unlikely to have symmetries matching those of the 
background galaxy groups, and because we only consider averages of the group satellite
ellipticities, we do not expect lensing distortions to bias our results.

\subsubsection{Weighted estimators for radial alignment measures}
\label{sec:weightedestimators}
To down-weight galaxies with noisy shape estimates and to better apply our results to predictions of 
weak lensing intrinsic alignment contamination, we compute the mean ellipticity components using an inverse
noise weighting per galaxy common for lensing measurements,
\begin{equation}
w_{e} \equiv \left(e_{\rm RMS}^2 + \sigma_{\rm e}^2\right)^{-1},
\end{equation}
where $\sigma_{\rm e}$ is the measurement error per ellipticity 
component and $e_{\rm RMS}$ is the r.m.s. ellipticity magnitude of our sample. 
Both the REGLENS and SIGMA samples have $e_{\rm RMS}\approx0.37$.
Because we are using bright galaxies, $e_{\rm RMS}$ is typically much larger than the ellipticity measurement errors yielding 
nearly equal weighting for most galaxies when computing the mean ellipticity components.

While the position angle is not a statistic used for lensing measurements, we adopt identical weights 
for computing the mean position angles of our samples in order to down-weight galaxies with large
position angle uncertainties. We derive position angle measurement uncertainties  
by formally propagating the ellipticity component measurement uncertainties. We therefore find a 
strong correlation between position angle error and ellipticity magnitude, with more round galaxies 
having larger position angle uncertainties. Our weights then favour more elliptical galaxies when
computing the mean position angles.

\subsubsection{Comparison of galaxy shape estimators}
\label{sec:shapeestimatorcomparison}
\begin{figure}
\centerline{
\includegraphics[width=0.5\textwidth]{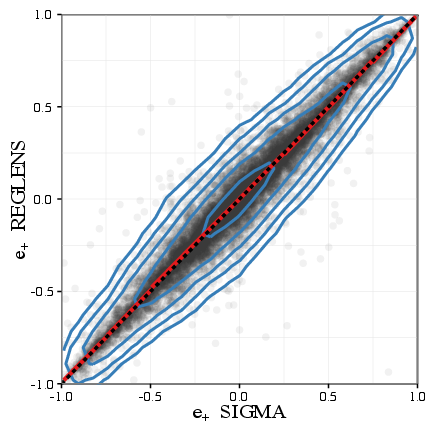}
}
\caption{\label{fig:eplusComparison}Comparison of the ellipticity component measuring radial 
alignment derived from two different galaxy shape estimators for all groups 
with three or more members. 
The logarithmically spaced blue con tours trace the density of the plotted points.
The dashed black line shows a slope of one going through the origin. 
}
\end{figure}

In Fig.~\ref{fig:eplusComparison} 
we compare the $\eplus$ components for all satellite galaxies in our group catalogue that 
passed the REGLENS galaxy shape quality cuts.
The dashed line in Fig.~\ref{fig:eplusComparison} has a slope of one and passes through the 
origin.
There is overall good agreement between the two shape estimators with larger scatter 
for smaller ellipticity magnitudes (i.e. rounder images) as expected.

\begin{figure}
\centerline{
\includegraphics[width=0.5\textwidth]{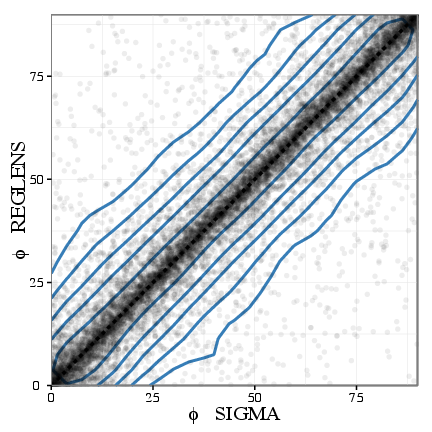}
}
\caption{\label{fig:posangleComparison} Comparison of the satellite galaxy position angle 
with respect to the group radius vector derived from two different galaxy 
shape estimators for all groups with three or more members.
The logarithmically spaced blue con tours show the density of points for all satellite galaxies with 
REGLENS $|e| > 0.05$.
The dashed black line shows a slope of one going through the origin. 
}
\end{figure}

We compare the position angle for our two shape estimators in Fig.~\ref{fig:posangleComparison}.
The scatter in the position angle derived from our two shape estimators 
is comparable to our estimate in the formal error on the mean position angle of 6$^{\circ}$, 
indicating that global statistical measures based on the two shape estimators should have 
a high degree of consistency.
By selecting only those galaxies with ellipticity magnitudes greater than $\sim 0.4$ we 
find we can reduce the scatter between the shape estimators seen in 
Fig.~\ref{fig:posangleComparison}, indicating some of the scatter is due to different 
handling of rounder galaxies where the position angle becomes poorly defined.

As previously pointed out by \citet{siverd09}, 
large isophote ellipticity 
does not necessarily imply high shape measurement accuracy in the presence of systematic errors such as 
isophotal twisting or confusion with nearby objects. 
Because SIGMA orientations are more sensitive to galaxy shapes at larger radii than 
REGLENS, isophotal twisting may contribute to the scatter in 
Fig.~\ref{fig:posangleComparison}. That is, physical differences in the shapes of galaxies 
at different galactic radii are expected to produce scatter in the position angles 
derived from different shape estimators, even in the absence of other sources of 
uncertainty.

\subsection{Mock catalogues}
\label{sec:mocks}

To assess the significance of our measured radial alignment statistics
we measure identical statistics in an ensemble of mock galaxy group catalogues 
described in detail in~\citet{robotham11}.
\citet{robotham11} constructed nine mock catalogues based on populating galaxies 
in the Millennium $N$-body dark matter simulation\footnote{http://www.mpa-garching.mpg.de/galform/millennium/}, 
matching many properties of the galaxies 
and groups to the data to infer the unobservable properties of the dark matter haloes
surrounding each group.
For each of the 9 mock catalogues, we make 25 different mock realizations of the 
satellite galaxy alignments by assigning each satellite galaxy in the mock a projected alignment 
angle drawn from a parametrized distribution. The distribution of mean radial alignment 
measures derived from the $9\times25$ mock realizations gives us a theoretical probability 
for assessing the significance of the observed mean radial alignment statistics.

While it would be possible in principle to assign satellite galaxy orientations
according to the 3D shapes and orientations of the satellite dark matter haloes in the 
mocks, we take a simpler approach in this paper that is more directly related to the 
observable projected galaxy shapes and assign the
projected satellite position angles, $\phi$, by drawing from a 
von~Mises distribution~\citep{circstats}, 
\begin{equation}\label{eq:vonmises}
  p(2\phi|\mu, \kappa) \equiv \frac{e^{\kappa\cos(2\phi - \mu)}}{2\pi I_0(\kappa)},
\end{equation}
with the same parameters $\mu \in [-\pi, \pi]$ and $\kappa>0$ for every mock satellite, where
$I_{0}$ is the zeroth order modified Bessel function of the first kind.
Unlike the commonly used Gaussian distribution, 
the von~Mises distribution (defined here with argument $2\phi$) has finite support on $[-\pi,\pi]$, 
limiting $-\pi/2 \le \phi \le \pi/2$. This is important when the variance of $\phi$ is large. 
In the limit of small standard deviation $\sigma$, the parameter $\kappa \sim 1 / \sigma^2$. 
When $\kappa=0$ the von~Mises distribution becomes the uniform distribution over the defined interval of support.
So, any constraint on $\kappa > 0$ constitutes a detection of non-uniformity in the 
position angle distribution.
The mock satellite ellipticity magnitudes are drawn from a fit to the distribution of
observed ellipticity magnitudes for a given multiplicity cut and, in some cases, binning 
in group mass and radius.

Given the 2-D positions of each galaxy in the GAMA mocks and the group membership, we compute the 
angle of the radius vector to each group satellite galaxy and then add the angle $\phi$ drawn from 
equation~(\ref{eq:vonmises}). Because this algorithm is sensitive to the group centre definition, 
we draw systematic group centre offsets from a distribution that fits the histograms 
for the ``Iter'' group centres in fig.~3 of~\citet{robotham11}.

\section{Results}
\label{sec:results}
\begin{figure}
\centerline{
\includegraphics[width=0.5\textwidth]{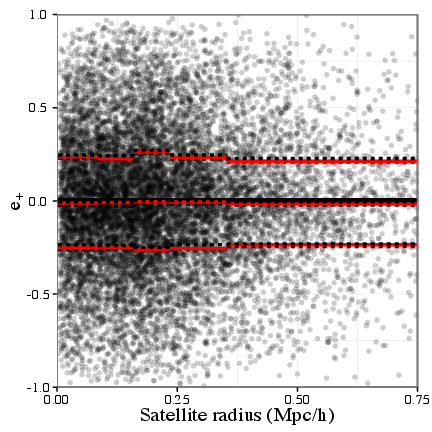}
}
\caption{\label{fig:eplusscatter}Ellipticity component measuring radial alignment 
derived from the REGLENS shapes 
for all groups with three or more members 
versus the projected physical distance from the group centre. There is no discernable 
dependence of the distribution of
ellipticity components on physical radius. The red lines show quartiles of the 
distribution of $\eplus$ components in radius bins. The black dashed lines show the 
error on the quartile measurements in each bin due to the finite number of galaxies.}
\end{figure}

The ellipticity component measuring radial or tangential alignment is shown 
in Fig.~\ref{fig:eplusscatter} versus projected physical 
separation from the group centre for all groups with 3 or 
more group members. We used the ``iterative centre'' 
and the median group redshifts from the GAMA group catalogue to calculate the 
projected physical separation of the satellites from the group centres.
The dashed red lines show the median and first and third quartiles of the 
$\eplus$ values in bins in the projected radius.
If the satellites in our catalogue were strongly radially aligned, the points 
and lines in Fig.~\ref{fig:eplusscatter} would be skewed below zero. 
Because the points and lines in Fig.~\ref{fig:eplusscatter} are approximately 
symmetric and broadly distributed about zero, we can infer
that any projected radial alignment signal 
in our group catalogue is sufficiently weak that we have limited statistical 
power to measure it.

For satellite orientations uniformly distributed in the plane of the sky,
the mean ellipticity components $\eplus$ and $\ecross$ 
should both be consistent with zero while the 
mean position angle should be consistent with 45$^{\circ}$.
On the other hand, for perfect radial alignments (i.e. $\phi=0$ for 
every galaxy in our catalogue) we would expect the mean $\eplus$ to be 
equal to the mean ellipticity of our catalogue,
$\left<e_{+}\right>\approx -0.46$, using the REGLENS shapes,
and $\left<e_{\times}\right> \approx 0$.
Note that Fig.~\ref{fig:eplusscatter} is not intended as an assessment 
of the detection or non-detection of radial alignments. Rather, we conclude from the 
symmetric distribution in 
Fig.~\ref{fig:eplusscatter} that measurements of mean radial alignment statistics should 
yield informative and useful summaries of the properties of the full statistical distribution.

\begin{figure*}
\centerline{
\includegraphics[width=0.5\textwidth]{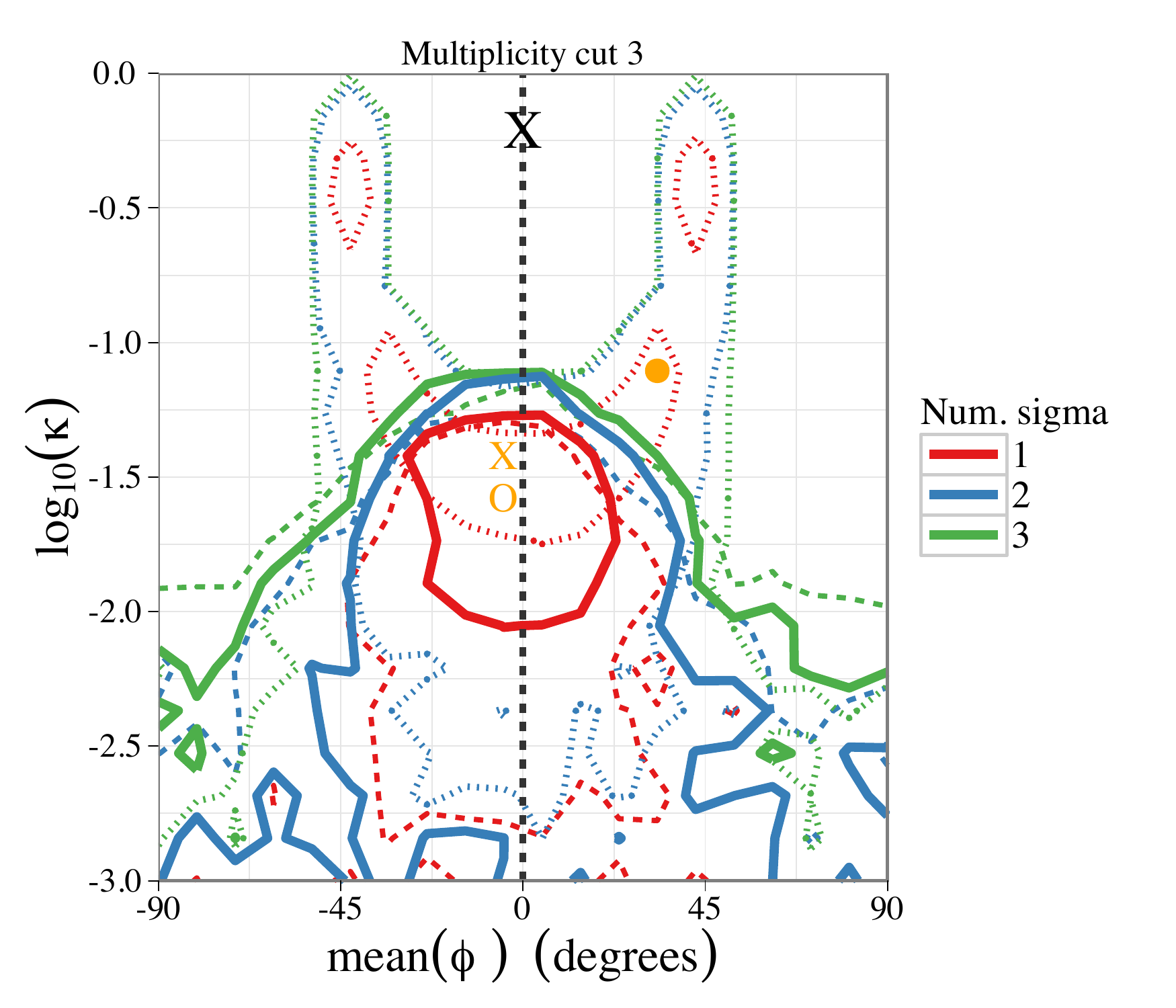}
\includegraphics[width=0.5\textwidth]{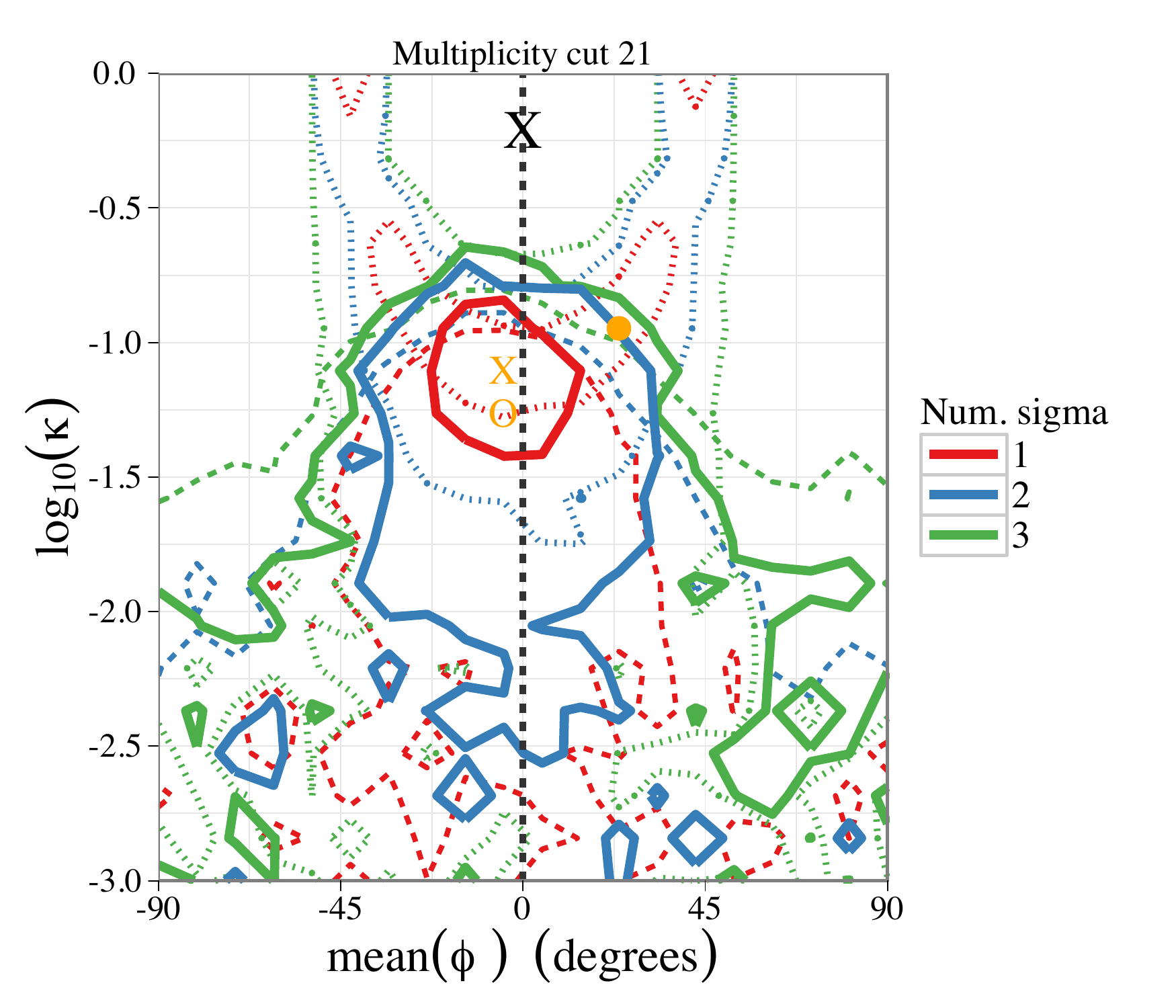}
}
\centerline{
\includegraphics[width=0.5\textwidth]{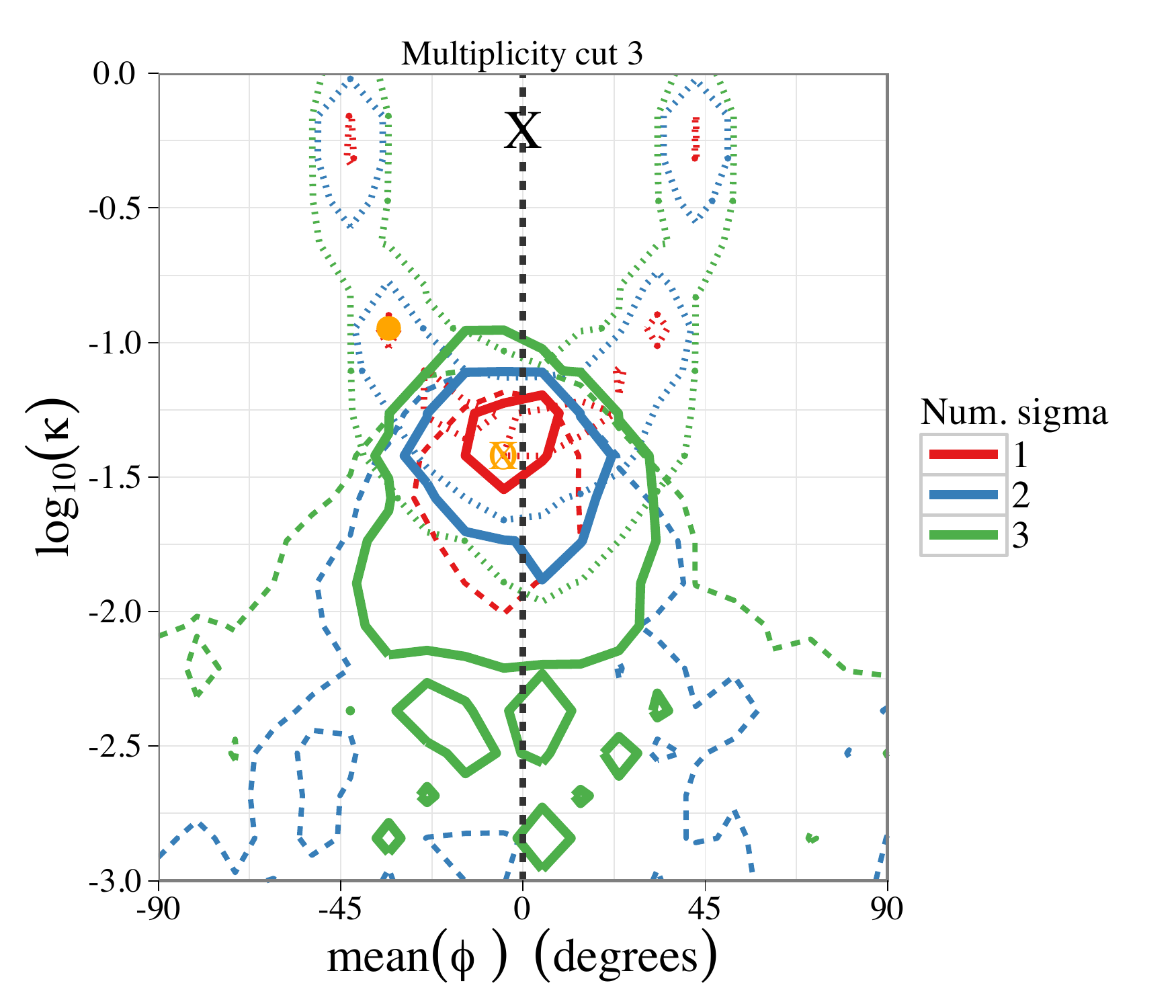}
\includegraphics[width=0.5\textwidth]{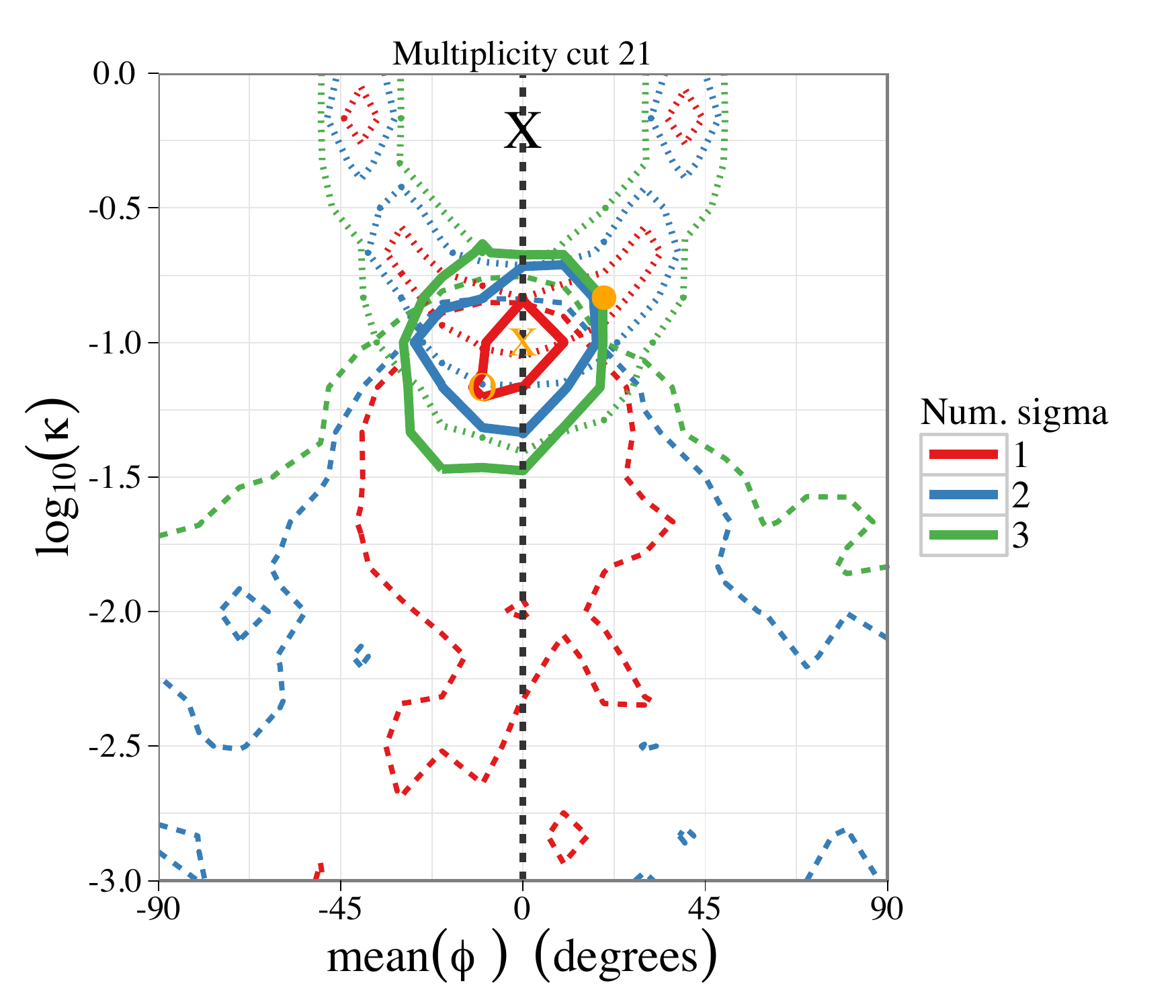}  
}
\caption{\label{fig:posteriorCon tours} Con tours of the posterior 
probability distribution for the 
two parameters, $\mu=\left<2\phi\right>$ and $\kappa$, of the
von~Mises distribution for $2\phi$ used to assign 
projected galaxy alignments in the 
mocks. The posterior is calculated given the combination of the 
mean ellipticity components and the mean position angles
for group multiplicity cuts of 3 (left-hand panel) and 21 (right-hand panel). 
The top panels use the REGLENS galaxy shape estimates while the bottom
panels use the SIGMA shape estimates.
The line styles denote posteriors for: mean ellipticity components 
($e_{+}$, $e_{\times}$, dotted), mean position angle mapped to $[0, \pi/2]$ (dashed), 
and the combination of both statistics including the cross-covariance (solid). 
Although the ellipticity components and position angles are perfectly correlated 
for a single galaxy, the ensemble means of these quantities include some independent 
information allowing us to obtain tighter constraints with the combination of 
statistics.
The orange open circle, filled circle, and ``X'' show the location of the  
maxima for each of these posteriors respectively. The black ``X'' near the top of 
each panel shows the model parameters for the cluster simulations shown by the 
short dashed green line in the right-hand panel of Fig.~\ref{fig:positionangle}.
}
\end{figure*}

As described in Section~\ref{sec:mocks}, 
we created 25 mock catalogue realizations for each of our nine mocks at 
each point in a grid of $\mu$ and $\kappa$ von~Mises distribution parameter values.
We then evaluated the posterior probability of the $\mu$ and $\kappa$ values 
at each grid point given the measured mean alignment statistics.
The 68\%, 95\% and 99\% (``1-3 sigma'') posterior con tours on $\mu$ and $\kappa$
are shown in Fig.~\ref{fig:posteriorCon tours} for group multiplicity cuts of 3 and 21. 
The dashed lines show the posterior con tours given the mean ellipticity components 
$\eplus$ and $\ecross$ while the dotted lines show the con tours for the mean position angle. 
Because we consider only the mean ellipticity components or position angles, which are 
not sufficient statistics for describing the full distribution of position angles in our 
catalogue, we can gain additional information by combining the mean ellipticity components 
and position angles using a covariance matrix derived from the ensemble of mock 
catalogue realizations. 
For individual satellite galaxies $\eplus$ and $\phi$ are 
perfectly correlated, but for our ensemble of mocks we typically find 
$\left<\eplus\right>$ and $\left<\phi\right>$ 
have a correlation coefficient $\sim 0.85$.
That is, the mean position angle cannot be derived from only the mean values 
of the ellipticity components and therefore contains some non-redundant information that can help 
further constrain the model for the alignment angle distribution.
The solid lines in Fig.~\ref{fig:posteriorCon tours} 
show the posterior con tours when the mean ellipticity components and  
position angle are jointly used to constrain the von~Mises distribution parameters.
The top panels use radial alignment statistics derived from the REGLENS galaxy shape estimates 
while the bottom panels use SIGMA-derived galaxy shapes.

The closed dashed and dotted red con tours in the top panels in 
Fig.~\ref{fig:posteriorCon tours} show that
the radial alignment signal is significant at no more than $\sim$1-$\sigma$
using the 
REGLENS galaxy shapes with either the ellipticity components or the position angle
radial alignment estimators. 
Combining the two estimators, groups with 21 or more members 
show a radial alignment signal at $\sim$2-$\sigma$ in the top right-hand panel of 
Fig.~\ref{fig:posteriorCon tours}.
The significance of the radial alignment detection increases to greater than 3-$\sigma$
when using
SIGMA to determine the galaxy shapes as shown in the bottom panels of 
Fig.~\ref{fig:posteriorCon tours}.
The black ``X'' near the top of each panel denotes parameter values that reproduce the simulated 
alignments in Fig.~10 of \citet{pereira08}. We rule out this model 
at more than 4-$\sigma$
with either of 
our galaxy shape estimators under the assumption that projected galaxy shapes perfectly trace 
the projected shapes of dark matter haloes in the simulations of~\citet{pereira08}.
The right-hand panel of Fig.~\ref{fig:positionangle} compares the best-fitting von~Mises probability 
distributions for $\phi$ in \citet{pereira08} (short-dashed green line) and from our data
using groups with three or more members and the REGLENS shapes (solid red line), corresponding 
to the orange cross in the top left-hand panel of Fig.~\ref{fig:posteriorCon tours}.

\begin{figure}
\centerline{
\includegraphics[width=0.5\textwidth]{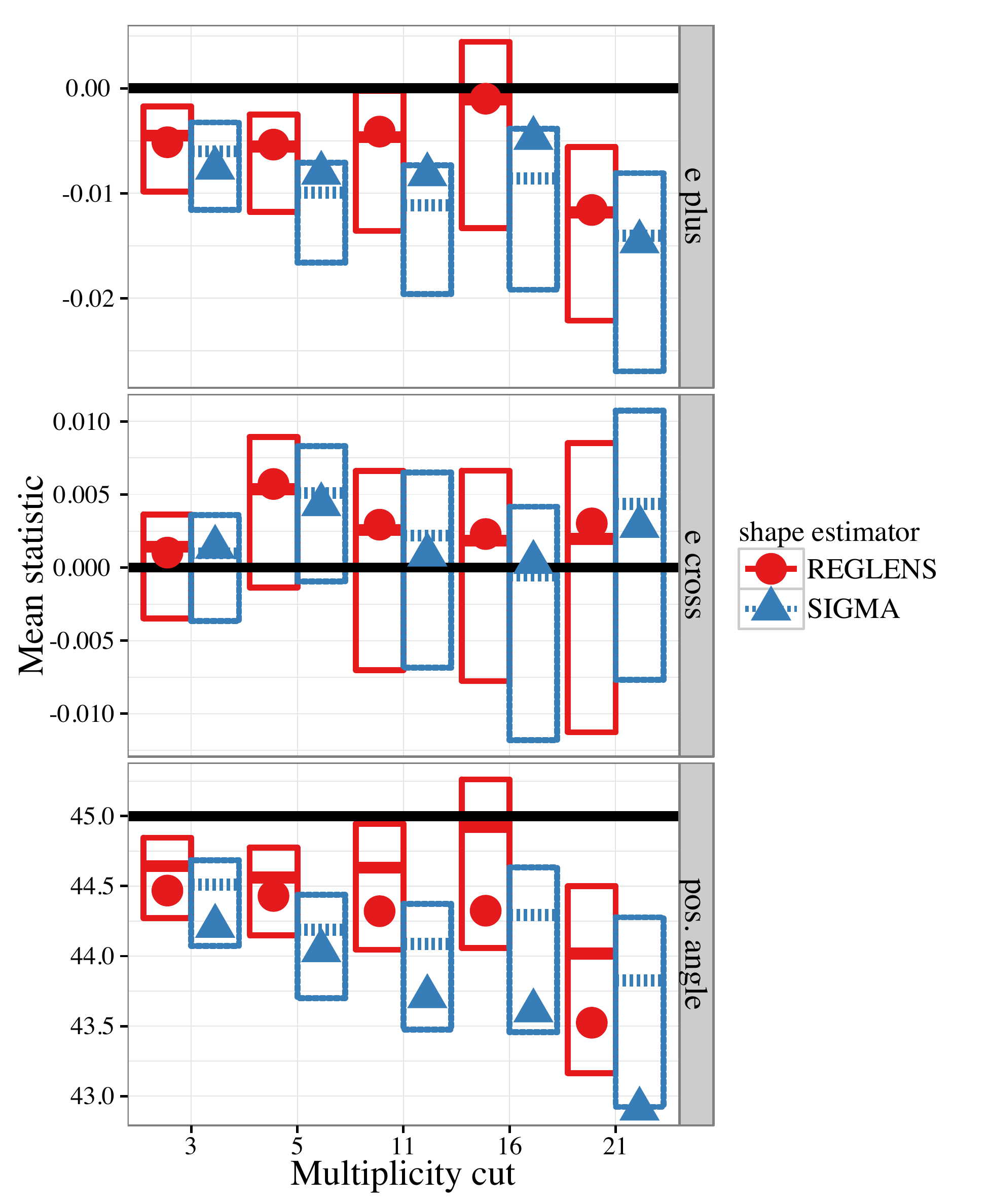}
}
\caption{\label{fig:globalMean}Mean ellipticity components and position angles in degrees as functions
of the minimum group multiplicity. 
The points show the mean observed values while the 
box ranges show the 68\% confidence intervals of the likelihoods given 
position angle distribution parameters chosen to maximize the 
conditional posterior given the observed mean ellipticities.}
\end{figure}

In Fig.~\ref{fig:globalMean} we show the mean alignment statistics for different 
group multiplicity cuts. For a given multiplicity cut, we determine confidence intervals 
on the radial alignment measures by first maximizing the posterior for the von~Mises 
distribution parameters $\mu$ and $\kappa$ and then finding the 68\% confidence intervals 
from the $25\times9$ mock realizations with specified $\mu$ and $\kappa$. 
We therefore quantify the uncertainty on the measured mean ellipticities and position angles 
using the width of the likelihood with fixed model parameters. This procedure is distinct from  
marginalizing the posterior for the von~Mises distribution parameters, which would yield larger
uncertainty intervals such that all our measurements would be consistent with a null signal.
Because our model for the projected radial alignment angles in the mocks is merely descriptive, rather 
than physically motivated, we believe our method of uncertainty quantification suffices for 
the current analysis. 

Our confidence intervals on the observed radial alignment measures
are shown by the boxes in Fig.~\ref{fig:globalMean}. The mean values from the mocks 
are shown by the horizontal lines in each box in Fig.~\ref{fig:globalMean}.
The circles and solid lines in Fig.~\ref{fig:globalMean}
show the results using the REGLENS shape estimates while 
the triangles and dashed lines show those for the SIGMA shape estimates. 
Note that the circles and triangles in Fig.~\ref{fig:globalMean} 
show observed mean values while the lines in the centres of the boxes 
show predicted median values values from the mocks. 
We include the 
mock-derived covariance between the mean ellipticity components and 
mean position angles when calculating the von~Mises distribution parameters 
that maximize the posterior.
Fig.~\ref{fig:globalMean} is one way to compare the effects of the group multiplicity cut
and the choice of galaxy shape measure on the significance of the radial alignment detection,
where we always choose the mock catalogue parameters that give the ``best fit'' 
to the observations (by maximizing the posterior). For a given galaxy shape estimator,
the mean alignment measures 
in Fig.~\ref{fig:globalMean} are consistent for different multiplicity cuts, with the 
exception of those groups with 21 or more members, which show slightly stronger 
radial alignments at 1-$\sigma$ significance.

We see in Fig.~\ref{fig:globalMean}
that the SIGMA shapes yield mean radial alignment measures systematically further from 
the null values than the REGLENS shapes. 
As described in, e.g.,~\citet{siverd09} and~\citet{hung12}, the 
position angle is potentially sensitive to systematic errors from 
poor angular resolution and close neighbours in the imaging.
While corrections are made in \citet{kelvin12} for such systematic errors,
we expect especially the SIGMA shapes
to remain contaminated at some level by close neighbours. 
Because the number density of satellite galaxies increases with decreasing 
projected group radius, it is possible that light bleeding from neighbouring galaxies could systematically
affect the observed $\eplus$ ellipticity component and position angle. Light bleeding from the BCG will
introduce a systematic bias as well increasing the detected radial alignment.
On the other hand, it is also possible that the outer shapes of galaxies as measured by 
SIGMA are more responsive to external tidal forces than the inner core of the galaxy.
We therefore continue to present the results from our two shape estimators together to assist later 
interpretation of these effects.

In Fig.~\ref{fig:globalMean} we always maximize the ($\mu$, $\kappa$) posterior given 
only the mean ellipticity components (i.e. neglecting the measured mean radial alignment angles),
which is denoted by the orange filled circle in each panel of Fig.~\ref{fig:posteriorCon tours}.
We found it difficult to find mock radial alignment parameters that simultaneously 
provide good fits to both the mean ellipticity components and the mean position angles 
for the multiplicity cuts of 16 and 21 in Fig.~\ref{fig:globalMean}. This could indicate 
that our mock radial alignment model is not sufficiently flexible to fit the data (i.e. 
more parameters are needed) or that there are numerous spurious position angle or 
ellipticity measurements in the high multiplicity groups. This model fitting choice 
is the reason that the points in the bottom panel of Fig.~\ref{fig:globalMean} are 
so far from the mock simulation mean values.

\begin{figure*}
\centerline{
\includegraphics[width=0.85\textwidth]{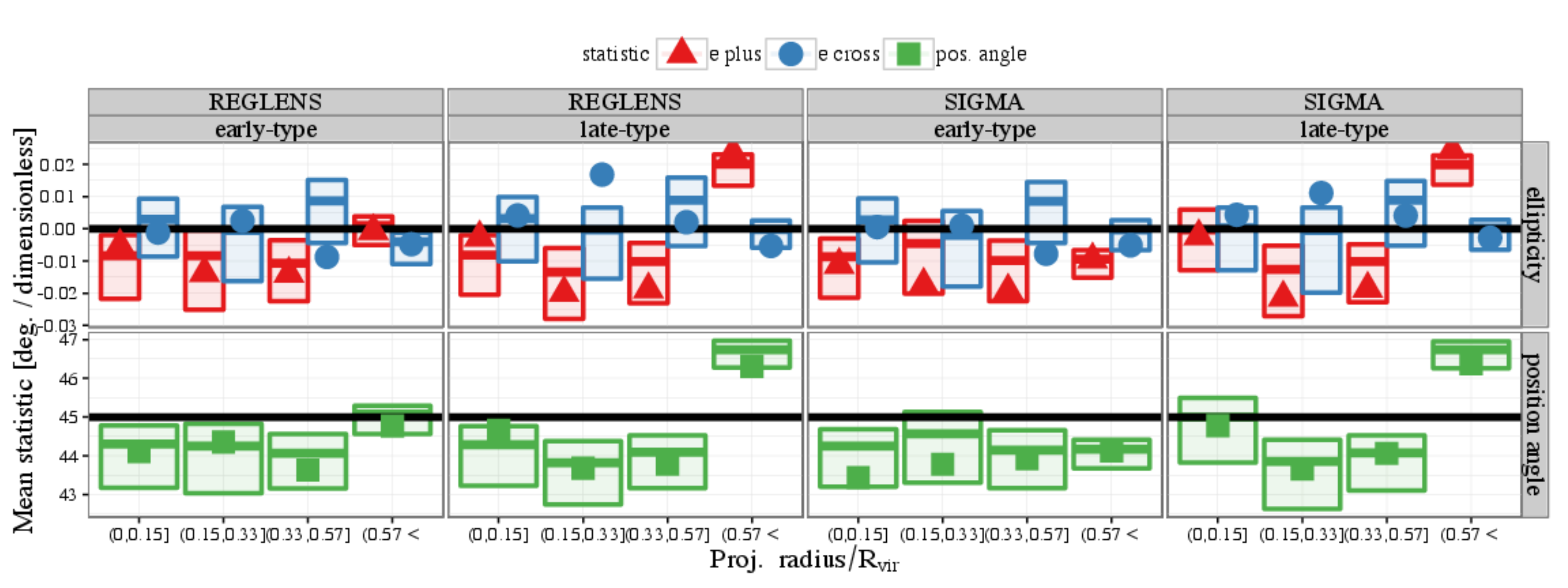}
}
\caption{\label{fig:radbinMeans} Mean ellipticity components and position 
angles similar to Fig.~\ref{fig:globalMean} but with multiplicity cut fixed at 3 and 
binned in normalized projected group radius. In contrast to Fig.~\ref{fig:globalMean}, 
the confidence intervals are chosen by maximizing the joint posterior for the position angle 
distribution parameters given both the observed mean ellipticity components and mean position angles, with 
each bin considered separately.}
\end{figure*}

\begin{figure*}
\centerline{
\includegraphics[width=0.85\textwidth]{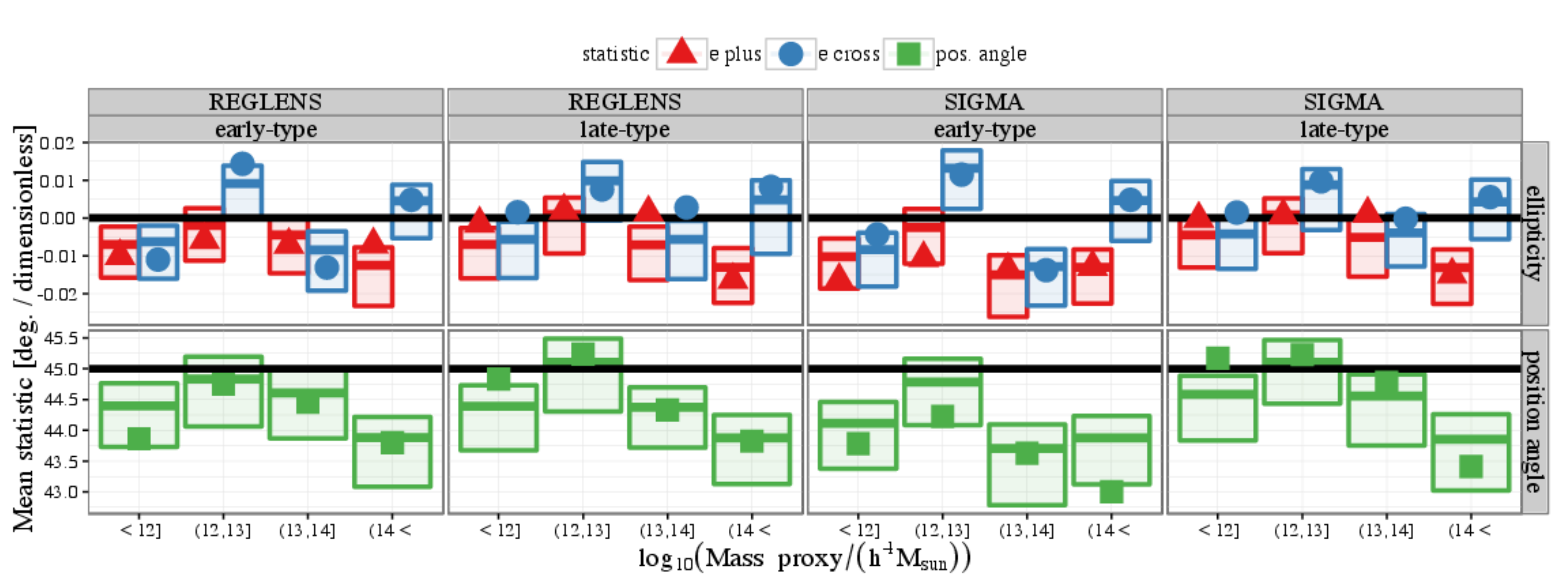}
}
\caption{\label{fig:massbinMeans}Mean ellipticity components and position 
angles similar to Fig.~\ref{fig:radbinMeans} but binned in the logarithm of the 
group mass proxy. 
The mass proxies are derived by matching the 
observed group sizes and velocity dispersions to those in the GAMA mock 
catalogues and taking matched dark matter halo masses~\citep{robotham11}, 
where the halo masses are those of the GALFORM Dhaloes~\citep{helly03} 
that are roughly equivalent to the enclosed mass equal to 200 times the critical density.}
\end{figure*}

\begin{figure*}
\centerline{
\includegraphics[width=0.85\textwidth]{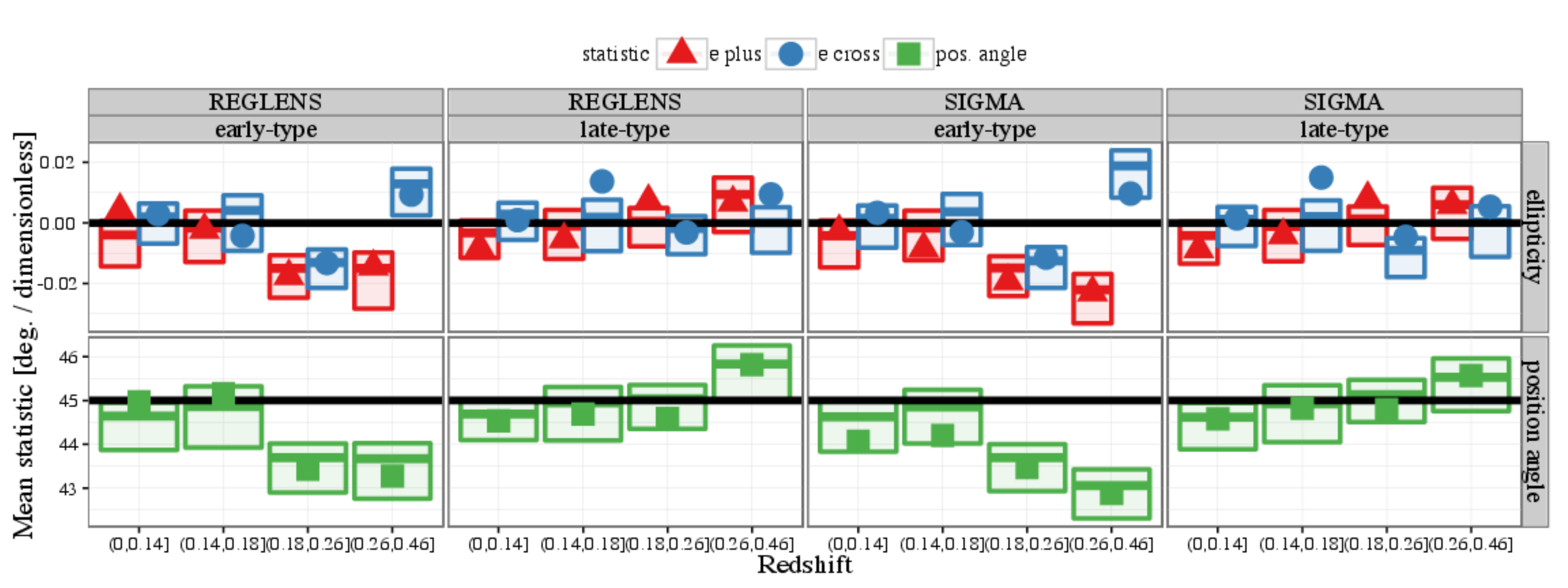}
}
\caption{\label{fig:zbinMeans}Mean ellipticity components and position 
angles similar to Fig.~\ref{fig:radbinMeans} binned in the median group redshift.}
\end{figure*}

To look for potential group mass, radius, or redshift dependence in the radial alignment signal, 
we recompute the mean radial alignment estimators in bins in the normalized group radius 
($r_p / \rvir$) in 
Fig.~\ref{fig:radbinMeans}, in group halo mass proxy~\citep[as described in][]{robotham11} 
in Fig.~\ref{fig:massbinMeans}, and in median group redshift in Fig.~\ref{fig:zbinMeans}. 
Because galaxies with different morphologies could be expected to respond to tidal torquing 
in different ways, we also split our galaxy sample in Figs.~\ref{fig:radbinMeans}, 
\ref{fig:massbinMeans}, and \ref{fig:zbinMeans} according to the galaxy S\'{e}rsic indices 
measured in~\citet{kelvin12}. We define galaxies with a S\'{e}rsic index greater than 2 
to be early-type and those with and index less than 2 to be late-type. This assigns 
48\% of our galaxy sample as early-type and 52\% late-type.

The shaded boxes in Figs.~\ref{fig:radbinMeans}, \ref{fig:massbinMeans}, and 
\ref{fig:zbinMeans} again 
show the 68\% confidence intervals on the mean radial alignment 
statistics derived from the mock realizations evaluated at the maximum 
posterior values for $\mu$ and $\kappa$, with the posterior maximized 
independently for each bin (rather than maximizing 
the joint posterior for all bin values simultaneously). Maximizing the 
posterior independently for each radius, mass, or redshift bin allows us to consider 
radial, mass, and redshift dependencies that are not explicitly modelled in the mocks.
That is, rather than defining a radial, mass, and/or redshift dependence 
for $\mu$ and $\kappa$ when generating mock realizations, we generate 
mocks with constant $\mu$ and $\kappa$ values and then cut the mocks in the 
same way as the observed catalogue to quantify the radial alignment significance 
in each cut sub-sample independently.
However, we cannot infer the joint significance of any radial, mass, or redshift 
dependence with our approach. Our measurements can identify important 
features needed in a radius-dependent model for the radial alignment angle 
distribution, but we are statistically limited by the size of our sample in 
constraining such a model.

Our radial alignment measures deviate from the expectation for random alignments 
at 1-$\sigma$ significance in
two of the three lowest radius bins for both shape estimators and 
galaxy types in Fig.~\ref{fig:radbinMeans}.
For the early-type galaxies, the largest radius bins in Fig.~\ref{fig:radbinMeans}
show radial alignment measures that are consistent with tangential alignment at 99\% confidence 
(or 3-$\sigma$). But, we are statistically limited in drawing any conclusions from 
the measured values at large radii.

The tidal torquing mechanism predicts stronger radial alignments 
at smaller fractions of the virial radius~\citep[except at radii where
satellite galaxies are at the perihelion of elliptical orbits,][]{pereira08}. 
Our uncertainties in Fig.~\ref{fig:radbinMeans} 
are too large to detect a radius dependence 
in the position angle distributions. We note again however, that 
our observed radial alignments are much weaker than that 
found for the alignments of haloes in $N$-body simulations~\citep{knebe08b}.
Our observed mean position angles are consistent at 2-$\sigma$ with previous measurements
using SDSS isophotal shapes~\citep{pereira05, faltenbacher07}, which found equivalent 
values of $\left<\phi\right> \sim$42--44$^{\circ}$ with $\left<\phi\right>$ increasing 
with increasing group radius. However we systematically favour mean position angles closer 
to $45^{\circ}$ than in previous SDSS isophotal measurements, which we attribute to the 
PSF correction in our two shape estimators as discussed further in 
Appendix~\ref{sec:isophoteComparison}.
The mean $\ecross$ ellipticity component is always consistent with zero in agreement with 
our expectations in the absence of dominant systematic errors in the shape measurements.

Because all mass bins in Fig.~\ref{fig:massbinMeans} are consistent within 
the uncertainties, we do not detect any host halo mass dependence in the radial 
alignment distributions. 
The highest group mass bin in Fig.~\ref{fig:massbinMeans}
($M > 10^{14}\, h^{-1}M_{\odot}$) has a non-null detection of radial alignments at 
68\% confidence for both galaxy types and shape estimators.
It is noteworthy in this context that \citet{knebe08,knebe08b} find that 
the subhalo radial alignment distributions in $N$-body simulations are independent 
of host halo mass.

Similarly in Fig.~\ref{fig:zbinMeans} we do not detect 
any redshift dependence in the radial alignment distributions, except 
for early-type galaxies using the SIGMA-derived shapes. In that case, 
the lowest and highest redshfit bins are inconsistent within their 68\% 
confidence intervals, with the radial alignments stronger at high redshift. 
The significant radial alignment in the two highest 
redshift bins for early-type 
galaxies in Fig.~\ref{fig:zbinMeans} 
(versus null detections in the two lowest redshift bins)
may be due to a correlation between halo mass proxy and redshift in our catalogue.
Within the highest mass bin in Fig.~\ref{fig:massbinMeans}, 40\% of the groups 
have redshifts greater than 0.25 and are therefore also in the highest 
redshift bin in Fig.~\ref{fig:zbinMeans}. But, our catalogue is too small
to explore this correlation in more detail.

\subsection{Lensing contamination} 
\label{sub:lensing_contamination}
Our measurement of $\left<\eplus\right>$ can be used to estimate the contamination of the 
weak lensing signal around galaxy groups when galaxies in the lens plane are 
confused with background source galaxies due to photometric redshift uncertainties.

The surface mass density contrast is often measured to estimate masses of isolated 
objects from weak lensing~\citep[e.g.][]{blazek12},
\begin{equation}
  \Delta \Sigma = \Sigma_c \left(\gamma_{+}^{\rm G} + \gamma_{+}^{\rm IA}\right),
\end{equation}
where $\gamma_{+}^{\rm G}$ is the tangential galaxy shear induced by lensing,
$\gamma_{+}^{\rm IA}$ is a spurious shear due to intrinsic alignments of lens-plane 
galaxies that are mixed with the source sample, and $\Sigma_c$ is the lensing critical 
surface mass density~\citep[e.g.][]{bartelmann01}.

The shear is related to the observed ellipticity by the shear responsivity, 
$\gamma \approx \left<e\right> / 2\mathcal{R}$. The amount of intrinsic alignment 
contamination also depends on the fraction of ``source'' galaxies that are 
actually at the redshift of the lens plane~\citep{fischer00}. This depends on the photometric 
redshift uncertainty and the projected radius from the centre of the group as the 
number density of lens-plane galaxies increases towards the group centre. 
The dependence of the source sample contamination on group radius is called 
the ``boost factor'' $B(r)$ in \citet{blazek12}, where $(B(r)-1)/B(r)$ is the 
fraction of lens-plane galaxies in the source sample.

Assuming $\mathcal{R}=0.87$, our observed $\eplus$ for groups 
with three or more members implies 
\begin{equation}\label{eq:deltasigma}
  \Delta \Sigma = -10 \pm 200 \times (B(r)-1)\qquad hM_{\odot} {\rm pc}^{-2},
\end{equation}
at a mean radius of the sample of $\sim300$~$h^{-1}$kpc, and
where the errors are dominated by the sample variance estimated from the mock group
catalogues and give the 1-$\sigma$ uncertainties. For haloes of massive 
early type galaxies and an assumed photometric redshift contamination 
at this radius of $B(r)-1=0.1$, the estimated $\Delta\Sigma$ in equation~(\ref{eq:deltasigma}) is 
approximately $-1\pm20$\% of the lensing signal measured in~\citet{mandelbaum06b}.

\section{Conclusions}
\label{sec:conclusions}
We have constrained the distribution of satellite galaxy radial alignment angles 
within GAMA groups independently as functions of group multiplicity, radius, 
mass proxy, and redshift.
For all subsets of our catalogue and definitions of galaxy shape estimators we consider, 
we observe a statistically weak radial alignment signal 
that is different from the predictions from dark matter $N$-body simulations.
Our comparisons with simulations are complicated 
by the fact that most published simulations make predictions
about (unobservable) dark matter alignments, rather than observable baryon alignments.
We conclude that our measurements give strong evidence for 
large misalignments between dark matter and baryonic (i.e. stellar) shapes, which imply
there are fundamental and important baryonic physical processes
that decouple the baryons from the dark matter in group and cluster environments~\citep[e.g.][]{sharma12}.
However, it is also possible that the dark matter may still be coupled with the baryons and have the
same alignments, in which case the predictions of the dark matter
alignments from simulations are incorrect. We consider this latter possibility unlikely, 
but mention it here for completeness.

The degree and significance of the radial alignment statistics depend on the 
method used to measure satellite galaxy shapes. 
Using PSF-corrected 2D model fits measured in SDSS imaging to
define satellite
galaxy orientations~\citep{kelvin12}, 
we detect satellite radial alignments at greater than 99\% confidence for all 
group multiplicities, but with mean position angles systematically larger 
than previous measurements in SDSS (that had no PSF corrections).
Using galaxy shape estimates optimized for weak lensing, 
we detect radial alignments at a weaker 95\% confidence but find 
best-fitting radial alignment angle distributions of similar width to those 
inferred from the SIGMA shapes. We use an ensemble of mock group catalogues based 
on $N$-body simulations to estimate the sample variance errors of our measurement, 
which are the dominant source of uncertainty.

For both our galaxy shape estimators, our 
non-uniform radial alignment detections are most significant 
at group radii less than $\sim$0.4~$\rvir$, at group masses larger than $\sim10^{14}h^{-1}M_{\odot}$, 
or at redshifts larger than $\sim 0.17$. But we do not have sufficient statistics 
to bin in combinations of these group properties. 
Also, our sample variance-dominated uncertainties are too large to detect any 
clear dependence of the radial alignments on group radius, mass or redshift.
Finally we note that radial alignment measurements at 
small radii and high redshifts are most likely to be susceptible to
systematics such as position angles errors at small radius 
and less accurate galaxy sizes and ellipticity measurements for small galaxies at high redshift.

While the trends in our data are consistent with the predictions from 
$N$-body simulations that find radial alignments to be 
created by tidal torquing within the group gravitational potential, our measured 
alignments using either of our galaxy shape estimators
are weaker than any existing predictions in the literature.
We speculated that the slightly larger radial alignments detected using the 
galaxy S\'{e}rsic model fits may indicate that tidal torquing acts to align the outer 
shapes of galaxies more efficiently, but we are limited in 
exploring this mechanism further by our sample size.

The radial alignments of satellite galaxies are also a concern for weak lensing 
measurements of group and cluster masses when lens and source galaxy samples must 
be inferred via photometric redshifts~\citep[e.g.][]{blazek12}. If the galaxies in the lens plane have strong 
intrinsic alignments and if some lens plane galaxies are mistaken for background sources, 
the lensing measurements can become biased. With our observed mean ellipticity components 
that include typical lensing inverse noise weights, we have discovered that intrinsic alignments
may be less than a 20\% contamination for photometric weak lensing measurements of high mass groups.
We leave a more thorough modelling of the intrinsic alignment contamination in lensing measurements 
for further work.
Our results also have implications for the magnitude of the small-scale 
intrinsic galaxy alignment contamination to cosmic shear measurements, where the 
predictions of~\citet{schneider10} are likely to be an overestimate of the small-scale 
cosmic shear contamination.

In future work we also plan to study the 3D shapes of
sub-haloes in the GAMA mocks to understand both what 3D misalignments are required to 
match the observations and how the projection of 3D triaxial galaxy shapes should 
be interpreted~\citep[see, e.g.][]{bett12}.

\section*{Acknowledgements}
We thank Jonathan Blazek for helpful feedback on an early version of this paper
and an anonymous referee for many helpful improvements including the suggestion
to measure alignments for different galaxy morphologies.
PN acknowledges a Royal Society URF and ERC StG grant (DEGAS-259586).
Part of this work performed under the auspices of the 
U.S. Department of Energy by Lawrence Livermore National Laboratory under Contract DE-AC52-07NA27344.

GAMA is a joint European-Australasian project based around a spectroscopic campaign using the 
Anglo-Australian Telescope. The GAMA input catalogue is based on data taken from the 
Sloan Digital Sky Survey and the UKIRT Infrared Deep Sky Survey. Complementary imaging of the 
GAMA regions is being obtained by a number of independent survey programs including {\it GALEX} MIS, 
VST KIDS, VISTA VIKING, {\it WISE}, Herschel-ATLAS, GMRT and ASKAP providing UV to radio coverage. 
GAMA is funded by the STFC (UK), the ARC (Australia), the AAO, and the participating institutions. 
The GAMA website is http://www.gama-survey.org/ .

Funding for the SDSS and SDSS-II has been provided by the Alfred P. Sloan Foundation, the Participating Institutions, the National Science Foundation, the U.S. Department of Energy, the National Aeronautics and Space Administration, the Japanese Monbukagakusho, the Max Planck Society, and the Higher Education Funding Council for England. The SDSS Web Site is http://www.sdss.org/.
The SDSS is managed by the Astrophysical Research Consortium for the Participating Institutions. The Participating Institutions are the American Museum of Natural History, Astrophysical Institute Potsdam, University of Basel, University of Cambridge, Case Western Reserve University, University of Chicago, Drexel University, Fermilab, the Institute for Advanced Study, the Japan Participation Group, Johns Hopkins University, the Joint Institute for Nuclear Astrophysics, the Kavli Institute for Particle Astrophysics and Cosmology, the Korean Scientist Group, the Chinese Academy of Sciences (LAMOST), Los Alamos National Laboratory, the Max-Planck-Institute for Astronomy (MPIA), the Max-Planck-Institute for Astrophysics (MPA), New Mexico State University, Ohio State University, University of Pittsburgh, University of Portsmouth, Princeton University, the United States Naval Observatory, and the University of Washington.

\bibliography{ia}

\appendix
\section{Galaxy isophotal shape comparison}
\label{sec:isophoteComparison}
In Section~\ref{sec:galaxyshapes} we introduced two methods for estimating 
galaxy shapes, the REGLENS pipeline optimized for weak lensing measurements 
and $r$-band S\'{e}rsic model fits (SIGMA). For the S\'{e}rsic fits, 
we use the output of the SIGMA pipeline developed for the GAMA survey and 
described in~\citet{kelvin12}. However many previous measurements of 
galaxy radial alignments in the literature have relied on isophote measurements 
provided in the SDSS {\tt CAS} catalogue. For the measurements we present here, 
a key difference between the SIGMA and {\tt CAS} pipelines is that SIGMA 
includes corrections for the rounding effect of the PSF that are not applied in the {\tt CAS} measurement. 
We also note that the {\tt CAS} catalogue does not provide any isophote measurement uncertainties.
In this section we compare some galaxy 
properties derived from the SDSS {\tt CAS} and GAMA SIGMA pipelines. 
For completeness, we also include results from the GAMA team IOTA pipeline that is based
on non-PSF corrected nine-band isophote measurements~\citep[][Liske, in preparation]{hill11}. 
Because 
of the lack of PSF correction, we expect the IOTA and {\tt CAS} measurements to be similar in 
yielding rounder galaxy shapes than the SIGMA and REGLENS pipelines.

\begin{figure*}
\centerline{
\includegraphics[width=0.55\textwidth]{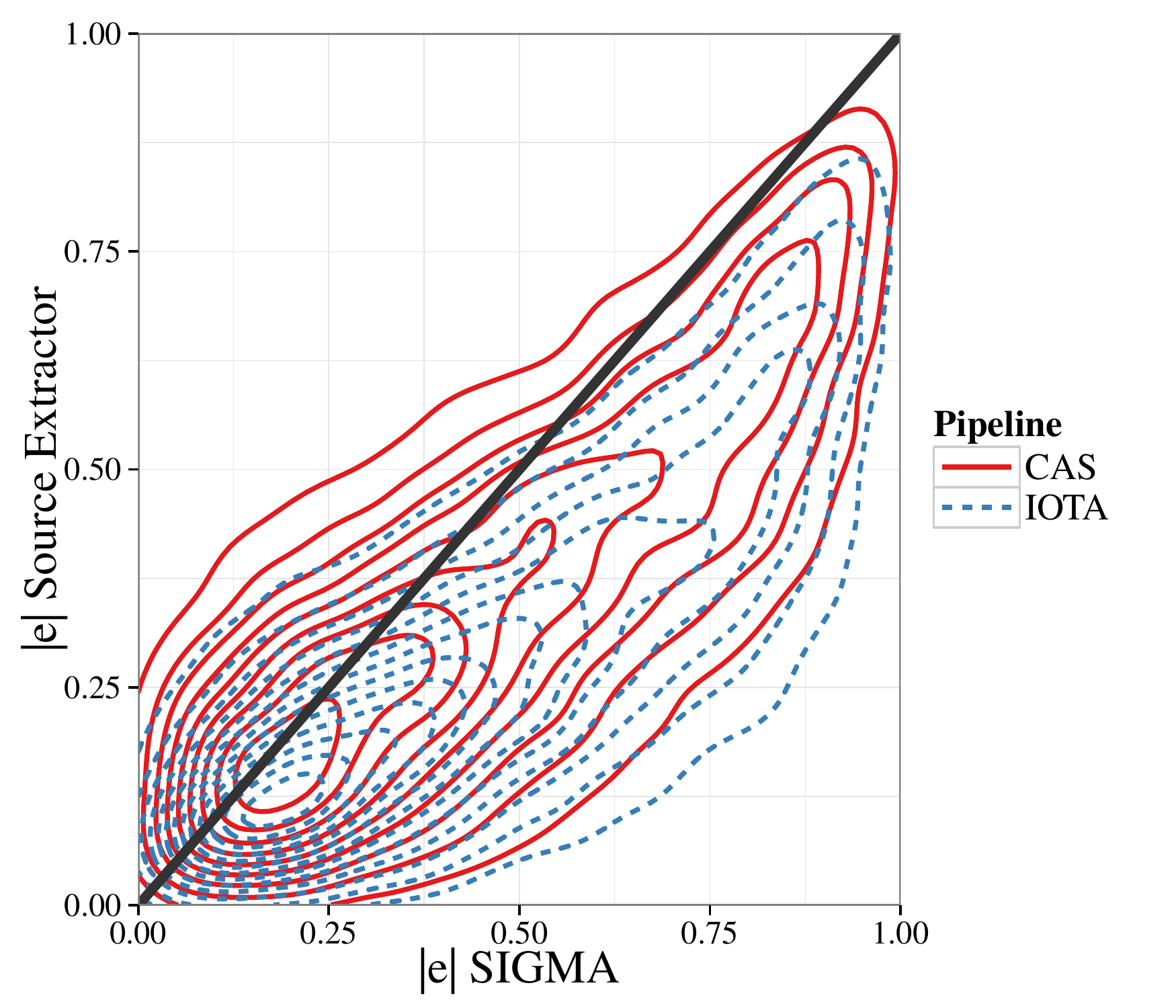}
\includegraphics[width=0.45\textwidth]{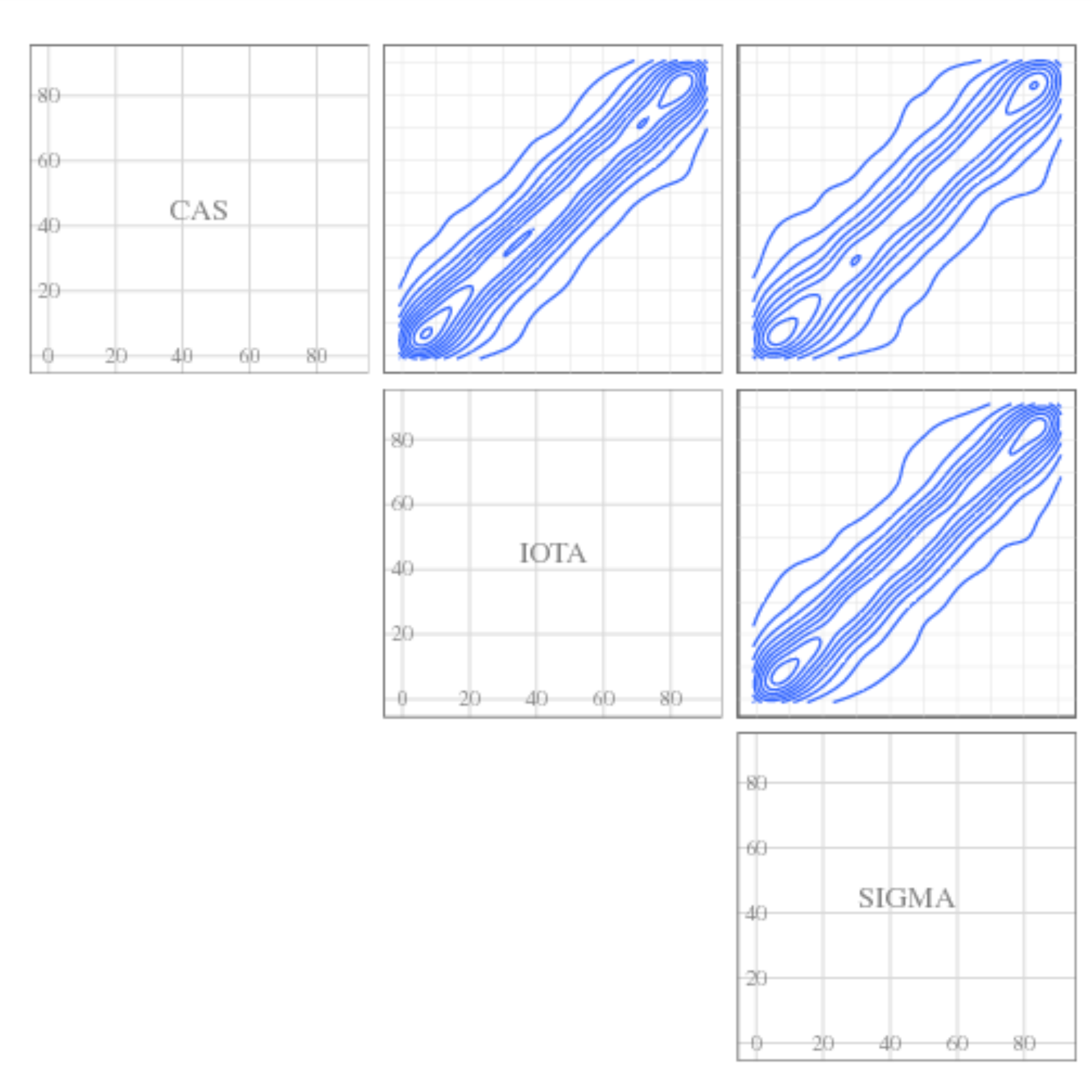}
}
\caption{\label{fig:isophoteComparison}Comparison of galaxy shape 
properties using the SDSS {\tt CAS} (isophotal) 
and GAMA SIGMA (model fit) and IOTA (isophotal) measurements 
(based on similar SDSS imaging data). The left-hand panel compares the 
ellipticity magnitudes while the right-hand panel compares the radial position 
angles with respect to the group centres.}
\end{figure*}
In the left-hand panel of Fig.~\ref{fig:isophoteComparison} we show the galaxy
ellipticity magnitudes derived from both the {\tt CAS} and IOTA isophote measurement pipelines 
compared with the SIGMA measurements.
As expected, the SIGMA ellipticities are systematically larger than those 
derived from the {\tt CAS} and IOTA catalogues with a nonlinear relationship between 
the bias and the ellipticity magnitude (round galaxies stay round after 
PSF convolution).

We compare the radial projected position angles (as defined in the 
left-hand panel of Fig.~\ref{fig:positionangle})
in the right-hand panel of Fig.~\ref{fig:isophoteComparison}.
There is much larger scatter between the position angle measurements than 
in Fig.~\ref{fig:posangleComparison}, including 25\% of all galaxies with position angle mismatches 
larger than 15$^{\circ}$. 
Restricting the comparison to galaxies with major axes spanning more than 12 
SDSS pixels and ellipticity magnitude greater than 0.2 
reduces the fraction of galaxies with position angle mismatches 
larger than 15$^{\circ}$ to 15\%. This trend is qualitatively consistent with 
the expectation that the PSF would mostly affect the observed orientations of 
rounder galaxies with smaller angular sizes.

\begin{figure*}
\centerline{
\includegraphics[width=0.5\textwidth]{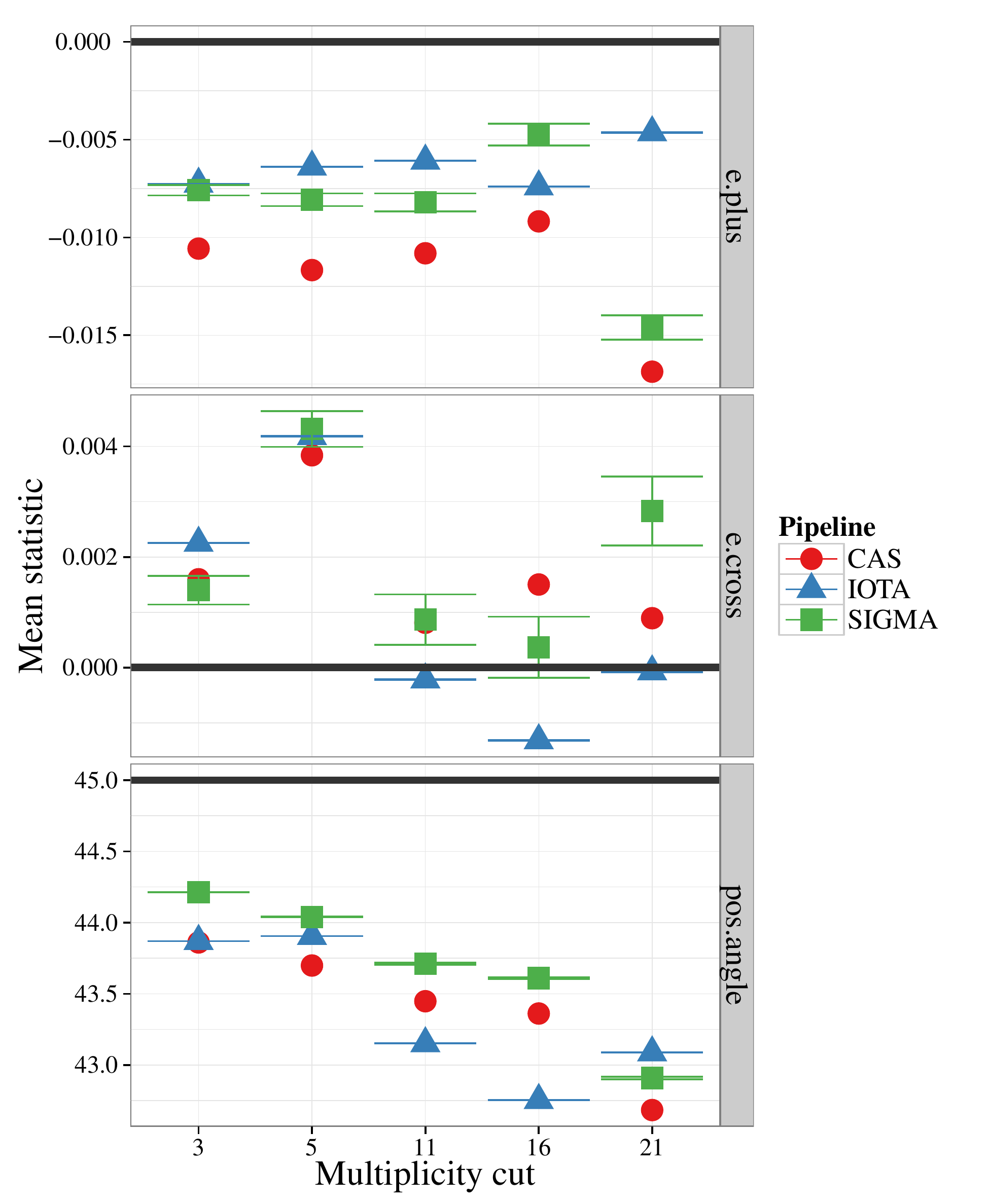}
}
\caption{\label{fig:isophotalMeanStats}Mean radial alignment statistics 
with different minimum group multiplicity cuts. The circles show the mean 
statistics derived when using the 25~mag per square arcsec. isophotes from the 
SDSS DR7 {\tt CAS} catalogue to estimate galaxy shapes and orientations. 
The triangles show analogous results when the galaxy isophotes are 
derived from the same SDSS imaging data but using the IOTA pipeline 
developed for the GAMA survey~\citep{kelvin12}.
The error bars for the IOTA measurements show the error on the mean statistics
derived by formally propagating the measurement errors on the isophote 
axis measurements for each galaxy. There are no such error estimates 
supplied in the {\tt CAS} catalogue. 
The squares show the SIGMA
derived values (based on S\'{e}rsic model 
fits), which are those used in the main body of the paper.
}
\end{figure*}
In Fig.~\ref{fig:isophotalMeanStats} we plot the mean radial alignment 
statistics as functions of minimum group multiplicity as in 
Fig.~\ref{fig:globalMean}. The error bars here show solely the measurement 
uncertainties, while the much larger sample variance errors are omitted.
For both the $\eplus$ and position angle 
statistics, the isophotal shapes from the {\tt CAS} catalogue yield systematically 
stronger radial alignment measurements than those from the SIGMA pipeline.
The mean position angles measured with the {\tt CAS} and IOTA catalogues 
show a less consistent trend however.
The $\ecross$ component in the middle panel of Fig.~\ref{fig:isophotalMeanStats}
(which is expected to be zero in the absence of systematics) has 
comparable magnitude for both isophote measurement pipelines.
We therefore conclude that the uncorrected effects of the PSF in the 
SDSS {\tt CAS} catalogue isophote measurements are likely to cause overestimates 
of the radial alignment of galaxies in groups and clusters. This 
is an important effect to consider when comparing our results with 
previous measurements relying on the SDSS catalogue.

\label{lastpage}
\end{document}